\documentclass{PoS}

\usepackage{graphicx}
\usepackage{xcolor}
\usepackage{tikz}
\usetikzlibrary{calc,shapes,arrows}
\usetikzlibrary{positioning,decorations.text,decorations.pathmorphing}

\tikzstyle{block} = [rectangle, draw, fill=blue!05, 
    text width=5em, text centered, rounded corners, minimum height=2.5ex]
\tikzstyle{wblock} = [rectangle, draw, fill=blue!05, 
    text width=6.em, text centered, rounded corners, minimum height=2.5ex]
\tikzstyle{sblock} = [rectangle, draw, fill=blue!10, 
    text width=3em, text centered, rounded corners, minimum height=2.5ex]
\tikzstyle{line} = [draw, -latex']

\pgfdeclaredecoration{gluon}{coil}
{
  \state{coil}[switch if less than=%
    0.5\pgfdecorationsegmentlength+%>
    \pgfdecorationsegmentaspect\pgfdecorationsegmentamplitude+%
    \pgfdecorationsegmentaspect\pgfdecorationsegmentamplitude to last,
               width=+\pgfdecorationsegmentlength]
  {
    \pgfpathcurveto
    {\pgfpoint@oncoil{0    }{ 0.555}{1}}
    {\pgfpoint@oncoil{0.445}{ 1    }{2}}
    {\pgfpoint@oncoil{1    }{ 1    }{3}}
    \pgfpathcurveto
    {\pgfpoint@oncoil{1.555}{ 1    }{4}}
    {\pgfpoint@oncoil{2    }{ 0.555}{5}}
    {\pgfpoint@oncoil{2    }{ 0    }{6}}
    \pgfpathcurveto
    {\pgfpoint@oncoil{2    }{-0.555}{7}}
    {\pgfpoint@oncoil{1.555}{-1    }{8}}
    {\pgfpoint@oncoil{1    }{-1    }{9}}
    \pgfpathcurveto
    {\pgfpoint@oncoil{0.445}{-1    }{10}}
    {\pgfpoint@oncoil{0    }{-0.555}{11}}
    {\pgfpoint@oncoil{0    }{ 0    }{12}}
  }
  \state{last}[next state=final]
  {
    \pgfpathcurveto
    {\pgfpoint@oncoil{0    }{ 0.555}{1}}
    {\pgfpoint@oncoil{0.445}{ 1    }{2}}
    {\pgfpoint@oncoil{1    }{ 1    }{3}}
    \pgfpathcurveto
    {\pgfpoint@oncoil{1.555}{ 1    }{4}}
    {\pgfpoint@oncoil{2    }{ 0.555}{5}}
    {\pgfpoint@oncoil{2    }{ 0    }{6}}
  }
  \state{final}{}
}
\tikzset{
    partial ellipse/.style args={#1:#2:#3}{
        insert path={+ (#1:#3) arc (#1:#2:#3)}
    }
}
\usepackage{amsmath}
\usepackage{amsfonts}
\usepackage{amssymb}
\usepackage{amsthm}
\usepackage{fixmath}
\usepackage{braket}
\usepackage{slashed}
\usepackage{fontenc}
\usepackage{dsfont}
\usepackage{bm}

\usepackage{lineno} 
\usepackage{fancybox}
\usepackage{adjustbox}
\cornersize*{5mm}
\linethickness{0.1mm}
\allowdisplaybreaks
\numberwithin{equation}{section}
\numberwithin{figure}{section}

\definecolor{red}      {rgb}{0.8,0.0,0.0}
\definecolor{green}    {rgb}{0.0,0.6,0.0}
\definecolor{darkblue} {rgb}{0.0,0.1,0.7}
\definecolor{brown}    {rgb}{0.6,0.1,0.0}
\definecolor{gray}     {rgb}{0.6,0.6,0.6}
\definecolor{darkgreen}{rgb}{0.0, 0.545098, 0.0}
\definecolor{orange}   {RGB}{238,80,25}
\definecolor{purple}   {rgb}{0.5,0.0,0.5}
\definecolor{babypink} {rgb}{0.64, 0.44, 0.44}

\renewcommand{\bm}{\mathbold}
\newcommand{\nn}{\nonumber}

\newcommand{\inv}{{^{-1}}}
\newcommand{\vt}{\vartheta}
\newcommand{\lt}{\lambda}

\newcommand{\oh}[1]{\widehat{#1}}

\newcommand{\sfs}{{\sf s}}
\newcommand{\sfx}{{\sf x}}
\newcommand{\sfy}{{\sf y}}
\newcommand{\sfz}{{\sf z}}

\newcommand{\ksxyz}{{\kappa_{\sfs,\sfx,\sfy,\sfz}}}
\newcommand{\ksyz}{{\kappa_{\sfs,\sfy,\sfz}}}

\newcommand{\ks}{{\kappa_{\sfs}}}
\newcommand{\gf}{{\gamma_5}}
\newcommand{\gt}{{\gamma_0}}
\newcommand{\gmu}[1]{{\gamma_{#1}}}
\newcommand{\ptau}[1]{{\oh P_\tau^{#1}}}

\newcommand{\pc}[1]{{\oh P_{\cal C}^{#1}}}

\newcommand{\abs}[1]{|#1|}

\newcommand{\nc}{N_c}
\renewcommand{\Re}{{\rm Re}\,}

\newcommand{\tr}{{\rm tr}\,}

\newcommand{\eq}[1]{\begin{equation}#1\end{equation}}
\newcommand{\al}[1]{\vskip-3ex\begin{align}#1\end{align}}
\newcommand{\bmat}{\left(\begin{array}}
\newcommand{\emat}{\end{array}\right)}

\title{QCD with Flavored Minimally Doubled Fermions}

\ShortTitle{QCD with Flavored Minimally Doubled Fermions}

\author{\speaker{Johannes Heinrich Weber}
\thanks{
J.H.W. was supported by the Excellence Cluster Universe, by the 
Bundesministerium f\"ur Bildung und Forschung (BMBF) under grant 
``Verbundprojekt 05P2015--ALICE at High Rate (BMBF-FSP 202) GEM-TPC Upgrade 
and Field theory based investigation'' and by the 
Kompetenznetzwerk f\"ur Wissenschaftliches H\"ochstleistungsrechnen in Bayern 
(KONWIHR) for the Multicore-Software-Initiative with the project 
``Production of gauge configurations at zero and non-zero temperature''.}
\thanks{
J.H.W. thanks M. Creutz and A. Kronfeld for very valuable discussions. 
}\\
        Technische Universit\"at M\"unchen, 
        Physik Department T30f, James-Franck-Str. 1, 
        85748 Garching, Germany\\
        E-mail: \email{johannes.weber@tum.de}}

\abstract{
I discuss minimally doubled fermions fermions as an ultra-local formulation 
on the lattice for sea quarks that realize a non-singlet chiral symmetry. 
I introduce a non-singlet mass term for Karsten-Wilczek fermions and 
identify the appropriate representation of the SU(2) flavor group at 
finite lattice spacing. 
I present an algebraic proof that the symmetry of the quark determinant 
under charge conjugation and reflections of the Euclidean axes is 
preserved for Karsten-Wilczek fermions as sea quarks.
Finally, I discuss how the flavor components in meson correlation functions 
with Karsten-Wilczek fermions emerge naturally and I show how taste-breaking 
can be avoided without fine tuning. 
}

\FullConference{34th annual International Symposium on Lattice Field Theory\\
		24-30 July 2016\\
		University of Southampton, UK}

\begin{document}

\vskip-1ex
\section{Introduction}
\vskip-1ex

Since the early years of lattice field theory the realization of 
chiral symmetry has always been an intricate topic. 
Minimally doubled fermions (MDF) are one of the few approaches that realize an 
exact chiral symmetry on the lattice. 
Hence, with MDF one may study light or even massless quarks. 
MDF are ultra-local and their Dirac operators are sparse matrices, so they are numerically faster than Ginsparg-Wilson fermions. 
Due to having a single pair of fermions, MDF have been considered for 
representing a degenerate light quark doublet. 
However, already the earliest studies suggested that hypercubic symmetry 
must be broken for MDF, giving rise to anisotropy, and thus, requiring extra
counterterms. 
Karsten's initial suggestion, its further refinement by Wilzcek, and the study 
of its broken charge conjugation and reflection symmetries by Pernici are the published works~\cite{kw} of the early era before the field fell dormant for a decade. 
The field was reinvigorated by Bori\c ci and Creutz in 2007, 
\mbox{cf. Refs.}~\cite{bc}, with many new contributions appearing in a 
brief time span thereafter. 
New patterns for attaining minimal doubling had been suggested though none 
of these made it beyond an exploratory stage. 
Only a few variants of MDF have been studied systematically with respect to 
their use in QCD. 
In particular, the symmetries of Karsten-Wilczek fermions (KWF) have been 
scrutinized again~\cite{bbtw}, their counterterms have been calculated 
perturbatively~\cite{ccww}, and the axial anomaly has been identified~\cite{tib}.

Numerical studies of KWF were published in \mbox{Refs.}~\cite{jhw}, 
where two independent non-perturbative methods for tuning the relevant 
counterterm have been identified, one based on the anisotropy of hadronic 
observables, the other based on a Fourier analysis of the changed pole 
structure of the quark propagators. 
It has been shown that the broken time reflection symmetry is usually 
not observed in meson correlation functions and an analytical proof has been 
presented. 
For KWF, Goldstone boson-like behavior of the pseudoscalar ground state 
at finite lattice spacing, the appearance of parity partner states, and taste 
breaking have been observed. 
Two higher-dimension operators necessary for the Symanzik improvement 
program have also been identified. 
However, previous numerical studies of MDF were performed in the quenched 
approximation. 
The chief worry~\cite{jhw} regarding dynamical MDF has been that it seems 
possible that the charge conjugation and reflection symmetries might be broken 
for the gauge configurations as well.
In the following, I will show that this concern is unfounded. 
In Section~\ref{sec:flavor}, I will introduce a non-singlet mass term for 
controlled flavor symmetry breaking and identify the appropriate 
representation of the flavor SU(2) group on the lattice. 
In Section~\ref{sec:determinant}, I present a proof that the vacuum does 
not suffer from the broken discrete symmetries. 
In Section~\ref{sec:mesons}, I outline the strategies for identifying 
flavored mesons and discuss how taste-breaking arises, and how it can be avoided. \vskip-2ex

\vskip-1ex
\enlargethispage{\baselineskip}
\section{Karsten-Wilczek fermions and flavor}
\label{sec:flavor}
\vskip-1ex

At tree level, degenerate KWF are realized with the Dirac operator
\eq{
 \mathcal D = D + m_0 + \left[3r K + r W\right],
 \qquad
 K = \frac{i}{a_\tau}\gt,
 \qquad
 W= -\frac{i}{a_\tau} \gt \left(3+a_\sigma^2 \bm D \cdot \bm D^* \right),
\label{eq:D}
}
where $D=\sum_\mu\gmu\mu D_\mu$ is the na\"ive Dirac operator. 
Together $K$ and $W$ form the Karsten-Wilczek term. 
The free propagator has only two poles at $k_0=0$ or $\pi/a$ (with $\bm k=0$) 
if $r^2>1/4$. 
The fermion action is invariant under the cubic group ($W_3$), parity 
($\mathcal P$), and joint application of charge conjugation ($\mathcal C$) 
and time reflection ($\mathcal T$), \mbox{Ref.}~\cite{bbtw}. 
Furthermore, $\mathcal D$ satisfies $\gf$ hermiticity.

Three counterterms are required in the interacting theory. 
The operators $c_3 K$ and $d_4\gt D_0$ serve as fermionic counterterms. 
In perturbation theory, $c_3$ is known to be an odd function of $r$. 
In the following, I incorporate the dimension four counterterm $d_4\gt D_0$ 
in the operator $D$, since it transforms in exactly the same way as long 
as no rotations beyond the cubic group $W_3$ are considered. 
Furthermore, I incorporate the coefficient $c_3$ in $\rho=3r+c_3$. 
The gluonic counterterm can be written in terms of temporal plaquette 
operators, \mbox{e.g.} the Wilson gauge action with counterterm reads 
\vskip-1ex
\eq{
 S^g[U] = \sum_n 
 \beta_\sigma \sum_{j<k} P^{jk}_n +
 \beta_\tau (1+d_p) \sum _{j} P^{j0}_n,
}
where $P^{\mu\nu}_m= \Re\tr (1-U^{\mu\nu}_m)/\nc$ is the normalized trace 
of the real part of the plaquette operator subtracted from its constant part. 
For the isotropic Wilson gauge action, the counterterms have been calculated 
at 1-loop level~\cite{ccww} and $c_3$ even has been determined 
non-perturbatively~\cite{jhw}. 
The global symmetries of the bare and the renormalized action are the same 
at finite lattice spacing.

A non-singlet mass term can be realized in the form\footnote{
There are various other possibilities to realize a non-singlet mass term, 
but unless a $W_3$ symmetric spatial hopping term is included, they will 
lead to an unfavorable flavor structure. 
} of 
\eq{
 m_3 M, 
 \qquad
 M=(1+a_\tau^2 D_0 D_0^*)(3+a_\sigma^2 \bm D \cdot \bm D^*),
\label{eq:M}
}
which leaves all of the previously mentioned symmetries intact. 
Of course, the chiral transform ($\chi=e^{i\gf\alpha}$) is a symmetry of 
the action only if both masses vanish ($m_0=m_3=0$). 
Nevertheless, there is another symmetry that is broken only by including 
the non-singlet mass term, \mbox{Eq.}~\eqref{eq:M}, in the Dirac operator, 
\mbox{Eq.}~\eqref{eq:D}. 
This is the so-called mirror fermion symmetry, which has been discovered 
by Pernici but generally been ignored. 
The mirror fermion symmetry arises because the action is invariant under 
joint application of time reflection or charge conjugation and a temporal 
shift transform ($\tau$). 
Here, I define the shift transform in terms of the self-inverse operator 
$\tau=\tau_0(-1)^{n_0}$, where $\tau_0=i\gt\gf$. 
While $M$ is invariant under $\mathcal C$ and $\mathcal T$, as required for 
a scalar mass operator, $M$ changes sign under $\tau$. 
Thus, $\tau$ is a representative of some SU(2) flavor matrix. 
The appropriate (self-inverse) representation of the full set of SU(2) flavor 
matrices is given by 
\eq{
 \bmat{c} \lt \\ \tau \\ \vt \emat = 
 \bmat{c} ~~~\gt (-1)^{n_\sigma} \\ i\gt\gf(-1)^{n_0} \\ ~~~\gf (-1)^{\bar n~} \emat =
 \bmat{c} \sigma_1 \times (-1)^{n_\sigma} \\ \sigma_2 \times (-1)^{n_0} \\ \sigma_3 \times (-1)^{\bar n~} \emat
 \otimes {\bf 1}_{2\times2},
\label{eq:su2}
}
where $n_\sigma=\sum_j n_j$ and $\bar n=\sum_\mu n_\mu$. 
$\{\lt,\tau,\vt\}$ trivially satisfy the SU(2) algebra.

\begin{table}
\vskip-1ex
\centering
\begin{tabular}{|c|c|c|c|c|c|c|c|c|}
 \hline
 $\mathcal O$ & $D$ & $K$ & $W$ & $\bf1$ & $M$ & $U^0$ & $U^j$ \\
 \hline
 $\chi$ &  $+D$ & $+K$ & $+W$ & $-\bf1$ & $-M$ & $U^0$ & $U^j$ \\
 \hline
 \hline
 $\lt$  &  $+D$ & $+K$ & $-W$ & $+\bf1$ & $-M$ & $U^0$ & $U^j$ \\
 $\tau$ &  $+D$ & $-K$ & $-W$ & $+\bf1$ & $-M$ & $U^0$ & $U^j$ \\
 $\vt$  &  $+D$ & $-K$ & $+W$ & $+\bf1$ & $+M$ & $U^0$ & $U^j$ \\
 \hline
 \hline
 $\mathcal P$ &  $+D$ & $+K$ & $+W$ & $+\bf1$ & $+M$ & $+U^0$ & $-U^{j\dagger}$ \\
 $\mathcal T$ &  $+D$ & $-K$ & $-W$ & $+\bf1$ & $+M$ & $-U^{0\dagger}$ & $U^j$ \\
 $\mathcal C$ &  $+D^T$ & $-K^T$ & $-W^T$ & $+\bf1^T$ & $+M^T$ & $+U^{0*}$ & $U^{j*}$ \\
 \hline
\end{tabular}
\caption{Symmetry pattern of the Dirac operator $\mathcal D_\kappa$. 
$d_4\gt D_0$ transforms like $D$ and is incorporated there.
}
\label{tab:symmetry}
\vskip-1ex
\end{table}

From these considerations, I derive the full symmetry pattern of the Dirac 
operator $\mathcal D_\kappa=\mathcal D+m_3M$ and summarize it in 
Table~\ref{tab:symmetry}. 
Due to the SU(2) algebra and the $\mathcal{C\!P\!T}$ invariance, out of each 
triplet (boxes in the table) only two transforms are independent. 
Hence, a set of five independent unitary transforms remains and each term in the Dirac operator $\mathcal D_\kappa$ transforms differently under the set. 
Only a single operator, namely $\mathcal P$, leaves each term invariant. 
Note that there is no unambiguous map between $\{\lt,\tau,\vt\}$ and the 
three SU(2) generators. 
However, in order to interpret $m_3 M$ in \mbox{Eq.}~\eqref{eq:M} as a 
flavor-diagonal mass term, the assignment to the generators should be 
as in the explicit $2\times2$ notation in \mbox{Eq.}~\eqref{eq:su2}: 
$\vt$ corresponds to $\sigma_3$. 
This suggests associating left-handed quarks with one flavor and 
right-handed quarks with the other.
Moreover, to avoid quarks with negative mass, the renormalized singlet and non-singlet masses must satisfy $m_0^R \geq m_3^R$. \vskip-2ex

\vskip-1ex
\section{Karsten-Wilczek fermions as sea quarks}
\label{sec:determinant}
\vskip-1ex

With KWF as sea quarks, the quark determinant is calculated from an operator 
$\mathcal D_\kappa$ that breaks four out of five independent discrete 
symmetries. 
In the following, I prove that the determinant nevertheless preserves all 
these symmetries. 
This outcome is analogous to twisted mass Wilson fermions, where the twist 
term breaks $\mathcal P$ while the vacuum is invariant under $\mathcal P$.

The key strategy is to introduce products of signs 
$\sfs,\sfx,\sfy,\sfz=\pm1$ in front of all operators that transform 
non-trivially under the broken symmetries, such that every sign corresponds 
to one of the four broken symmetries. 
Then no operator that depends explicitly on any of these signs is invariant,  
but any operator that is independent of all signs is invariant under 
all broken symmetries. 
With assignments $\sfs\leftrightarrow\mathcal C$, $\sfx\leftrightarrow\chi$, 
$\sfy\leftrightarrow\lt$ and $\sfz\leftrightarrow\vt$ 
I rewrite the full Dirac operator\footnote{$\kappa=(m_0,m_3,r,c_3,d_4;d_p,\beta_\sigma,\beta_\tau)$ 
is to be understood as a multi-index of bare parameters and counterterms. 
I use an extended multi-index $\ksxyz =(\kappa,\sfs,\sfx,\sfy,\sfz)$, 
since the role of $\sfs,\sfx,\sfy,\sfz$ is different from the `physical' 
parameters of $\kappa$. 
} 
as 
\eq{
 \mathcal D_\ksxyz = 
 D + \sfx m_0 +\sfs\sfz \rho K +\sfs\sfy r W  +\sfx\sfy m_3 M
\label{eq:Dk}
}
and find the eigenvalue equation for eigenvalue $\omega_\ksxyz$ and eigenvectors $\phi^\omega_\ksxyz$ to be 
\eq{
 \bar\phi^\omega_\ksxyz \mathcal D_\ksxyz \phi^\omega_\ksxyz=
 \omega_\ksxyz.
\label{eq:eve}
}
Next, I prove that $\det\mathcal D_\ksxyz$ is not affected by the 
broken discrete symmetries as it is independent of the signs 
$\sfs,\sfx,\sfy,\sfz$ and hence satisfies 
$\det\mathcal D_\kappa\equiv\det\mathcal D_\ksxyz$. 
In the following, I omit $\sfs,\sfx,\sfy,\sfz$ unless I 
specifically discuss the dependence on a sign and attach it to the multi-index 
$\kappa$, \mbox{e.g.} $\kappa_{+\sfx}$ or $\kappa_{-\sfx}$.

\vskip-1ex
\begin{enumerate}
\item[0)] 
If there are massless quarks (due to $m_0=m_3=0$ or due to cancellation 
between the two mass terms), then there is at least one exact zero mode and 
the determinant vanishes as it is the product of 
the eigenvalues of the Dirac operator, 
$\det\mathcal D_\kappa=\prod_{\omega_\kappa} \omega_\kappa.$
Thus, $\det\mathcal D_\kappa$ (being zero) would be trivially independent 
of all signs\footnote{
The arguments 2) and 3), which show that $\omega_\kappa\equiv\omega_{\ksyz}$ is 
independent of $\sfs$, $\sfy$ and $\sfz$, are independent of the masses.   
}. 
Hence, I exclude zero mass in the following. 
\item[1)] 
First, I use $\gf$ hermiticity of $\mathcal D_\kappa$. 
It follows that eigenvalues form complex conjugate pairs, so if 
$\omega_\kappa$ is an eigenvalue, so is $\omega_\kappa^*$. 
Thus, $\det\mathcal D_\kappa$ is real and strictly positive (since quark 
mass zero has been excluded). 
Next, I combine the positivity with chiral symmetry and obtain 
\vskip-2ex
\eq{
 0 < \det \mathcal D_{\kappa_{+\sfx}} =
 \sqrt{( \det \mathcal D_{\kappa_{+\sfx}} \det [\gf \mathcal D_{\kappa_{+\sfx}} \gf])} =
 \sqrt{( \det \mathcal D_{\kappa_{+\sfx}} \det [-\mathcal D_{\kappa_{-\sfx}}])}.
}
Since quark masses are non-zero, the mass matrix $m_0+m_3M$ is invertible and I find 
\vskip-2ex
\eq{
 0 < \det \mathcal D_{\kappa_{+\sfx}} =
 \sqrt{( \det (m_0+m_3M)^2 \det [1-\left(\frac{D+\rho K+rW}{m_0+m_3M}\right)^2])},
}
which is clearly independent of $\sfx$. 
In fact, even
$\abs{\omega_\kappa}\equiv\abs{\omega_{\kappa_{+\sfx}}}
=\abs{\omega_{\kappa_{-\sfx}}}$ 
is satisfied by each eigenvalue and $\det \mathcal D_\kappa$ can be written 
as 
$
 \det\mathcal D_{\kappa_\sfx}=
 \prod_{\omega_{\kappa_\sfx}}\omega_{\kappa_\sfx}=
 \prod_{\omega_{\kappa_\sfx}}\abs{\omega_{\kappa_\sfx}}=
 \prod_{\omega_\kappa}\abs{\omega_\kappa}.
$
\item[2)] 
Second, I plug the SU(2) generators $\lt$ or $\vt$ into the eigenvalue 
equation, \mbox{Eq.}~\eqref{eq:eve}. 
For $\lt$, I define 
$\varphi^\omega_{\kappa_{-\sfy}}=\lt\phi^\omega_{\kappa_{+\sfy}}$ and
$\bar\varphi^\omega_{\kappa_{-\sfy}}=\bar\phi^\omega_{\kappa_{+\sfy}}\lt$ 
and find an equality of eigenvalues, \vskip-2ex
\eq{
 \omega_{\kappa_{+\sfy}}=
 \bar\phi^\omega_{\kappa_{+\sfy}}
 \mathcal D_{\kappa_{+\sfy}}
 \phi^\omega_{\kappa_{+\sfy}}=
  \bar\phi^\omega_{\kappa_{+\sfy}}\lt\
 \lt\mathcal D_{\kappa_{+\sfy}}\lt\
 \lt\phi^\omega_{\kappa_{+\sfy}}=
 \bar\varphi^\omega_{\kappa_{-\sfy}}
 \mathcal D_{\kappa_{-\sfy}}
 \varphi^\omega_{\kappa_{-\sfy}}=
 \omega_{\kappa_{-\sfy}}.
}
Hence, $\omega_\kappa\equiv\omega_{\kappa_{+\sfy}}=\omega_{\kappa_{-\sfy}}$.  
Since $\sfy^2=1$ this implies that $\omega_\kappa$ is independent of 
$\sfy$. 
Using $\vt$, I reiterate the same steps and find that 
$\omega_\kappa\equiv\omega_{\kappa_{+\sfz}}=\omega_{\kappa_{-\sfz}}$ 
is also independent of $\sfz$. 
\item[3)] 
In the third step,I repeat this procedure by plugging the charge conjugation 
operator $\mathcal C$ into \mbox{Eq.}~\eqref{eq:eve}. 
I define charge conjugated eigenvectors 
$\left(\varphi^{\omega\,c}_{\kappa_{-\sfs}}\right)^T=-\bar\phi^\omega_{\kappa_{+\sfs}}\mathcal C$ and
$\left(\bar\varphi^{\omega\,c}_{\kappa_{-\sfs}}\right)^T=\mathcal C\phi^\omega_{\kappa_{\sfs}}$,
and find, \vskip-2ex
\al{
 \omega_{\kappa_{+\sfs}} &=
 \bar\phi^\omega_{\kappa_{+\sfs}}[U]
 \mathcal D_{\kappa_{+\sfs}}[U]
 \phi^\omega_{\kappa_{+\sfs}}[U]=
  \bar\phi^\omega_{\kappa_{+\sfs}}[U]\mathcal C\
 \mathcal C\mathcal D_{\kappa_{+\sfs}}[U]\mathcal C\
 \mathcal C\phi^\omega_{\kappa_{+\sfs}}[U] \nn\\
 &=
 -\left(\varphi^{\omega\,c}_{\kappa_{-\sfs}}[U]\right)^T
 \left(\mathcal D_{\kappa_{-\sfs}}[U]\right)^T
 \left(\bar\varphi^\omega_{\kappa_{-\sfs}}[U]\right)^T=
 \left(\bar\phi^\omega_{\kappa_{-\sfs}}[U]
 \mathcal D_{\kappa_{-\sfs}}[U]
 \phi^\omega_{\kappa_{-\sfs}}[U]\right)^T=
 \omega_{\kappa_{-\sfs}}.
}
Hence, $\omega_\kappa\equiv\omega_{\kappa_{+\sfs}}=\omega_{\kappa_{-\sfs}}$ 
is independent of $\sfs$ following the previous line of reasoning. 
\end{enumerate}
\vskip-3ex
Hence, I find that the moduli of the eigenvalues and the full determinant 
are independent of the signs, $\abs{\omega_\kappa}\equiv\abs{\omega_\ksxyz}$ 
and $\det\mathcal D_\kappa\equiv\det\mathcal D_\ksxyz$. 
Therefore, the quark determinant is not affected by the broken charge 
conjugation and reflection symmetries. 
This statement holds non-perturbatively and is independent of any parameters 
of the theory as well as of the gauge fields. 
Since the Wilson gauge action with counterterm is invariant under these 
symmetries, this also holds for the QCD vacuum/medium.
I note in passing that field configurations are automatically 
$\mathcal O(a)$ improved and depend on the bare parameters of the quark 
action only through $m_0^2,m_3^2$ and $r^2$. \vskip-2ex

\vskip-1ex
\section{Flavor components in meson correlation functions}
\label{sec:mesons}
\vskip-1ex

In the following I stress only the dependence on $\sfs$ and $m_3$, which 
encode taste and flavor structure of meson correlation functions. 
It is key to introduce charge conjugation and su(2) projectors, 
\mbox{i.e.} 
\eq{
 \pc\pm=\frac12\left(1\pm\mathcal C\right), 
 \qquad
 \ptau\pm=\frac12\left(1\pm\tau\right), 
}
and decompose $\mathcal D_\ks$ and the quark propagator $S=D_\ks^{-1}$ using the projectors. 
I find \vskip-2ex
\al{
 S&=S^E+S^O, &\quad
 S^E&=\left(E-OE\inv O\right)\inv, &\quad
 S^O&=-S^EOE\inv, 
\label{eq:SEO} \\
 S&=S^\eta+\sfs S^\omega, &\quad
 S^\eta&=\left(\eta-\omega\eta\inv \omega\right)\inv, &\quad
 S^\omega&=-S^\eta\omega\eta\inv 
\label{eq:Seo}
}
where $E=D+m_0$ and $O=m_3M+\sfs\rho K+\sfs rW$ as well as 
$\eta=D+m_0+m_3M$ and $\omega=\rho K+rW$ are the even and odd operators 
under $\ptau\pm$ and $\pc\pm$. 
Note that $\sfs$ is taken out from $\omega$ and $S^\omega$ explicitly.

Mesons are created and annihilated through quark-bilinear interpolating 
operators. 
The kernels $\Gamma$ of such interpolating operators at source ($\Gamma_a$) 
and sink ($\Gamma_b$) can be classified into even and odd matrices under 
$\pc\pm$ and $\ptau\pm$, \mbox{e.g.} for on-site operators under $\pc\pm$ 
the sets are  
\eq{
 \{\Gamma^\eta\}=\{{\bf 1},\gf,i\gmu\mu\gf\}, \quad 
 \{\Gamma^\omega\}=\{\gmu\mu,i\gmu\mu\gmu\nu\}.
\label{eq:Gamma}
}
Depending on the kernels used, only a limited set of the propagator 
components contribute to any given meson correlation function. 
In fact, for sufficiently symmetric choices (either 
$\Gamma_a,\Gamma_b\in\{\Gamma^\eta\}$ or 
$\Gamma_a,\Gamma_b\in\{\Gamma^\omega\}$) any terms linear in $\sfs$ that would violate 
$\mathcal C$ and $\mathcal T$ automatically vanish. 
One would exclusively have such unphysical terms for sufficiently 
antisymmetric $\Gamma_a,\Gamma_b$. 
These terms cancel explicitly upon averaging $\sfs=\pm1$ for the valence 
quarks. 
Because the quark sea is the same for $\sfs=\pm1$, this average does not 
introduce any partial quenching. 
I note in passing that such meson correlation functions and derived 
observables are automatically $\mathcal O(a)$ improved. 
In the following I always assume sufficiently symmetric $\Gamma_a,\Gamma_b$.

Next, I nest the projections of the propagator and take out any linear 
$m_3$ dependence, 
\eq{
 S=S^{\eta E}+m_3S^{\eta O} + \sfs m_3 S^{\omega E} +\sfs S^{\omega O}. 
\label{eq:S}
}
The kernels $\{\Gamma^{\eta E}\}$ and so on are obtained through an 
obvious generalization of \mbox{Eq}~\eqref{eq:Gamma}. 
The propagators and interpolator kernels are intertwined in the 
quark-disconnected contribution, \vskip-2ex
\al{
 C^{\rm disc}_{\Gamma_a,\Gamma_b}(n_0) &=\phantom{\big(} 
 \Braket{S^{\eta E}\Gamma^{\eta E}}\Braket{S^{\eta E}\Gamma^{\eta E}} +
 \Braket{S^{\omega O}\Gamma^{\omega O}}\Braket{S^{\omega O}\Gamma^{\omega O}} \nn\\
 &+\left(
 \Braket{S^{\eta O}\Gamma^{\eta O}}\Braket{S^{\eta O}\Gamma^{\eta O}}+
 \Braket{S^{\omega E}\Gamma^{\omega E}}\Braket{S^{\omega E}\Gamma^{\omega E}}
 \right)\times m_3^2 ,
}
where the $\Braket{\ldots}$ denotes combined color-spin trace and gauge 
average. 
Spatial summation is implied. 
Here, association of the traces with either source or sink is not relevant. 
Thus, the quark-disconnected contribution for 
$\gf=\Gamma_{a,b}\in\{\Gamma^{\eta O}\}$ 
starts at $\mathcal O(m_3^2)$, 
which unambiguously identifies the lowest-lying (non-vacuum) state as a 
neutral pion. 
I also find that $i\gt\gf=\Gamma_{a,b}\in\{\Gamma^{\eta E}\}$ starts at 
$\mathcal O(1)$, which identifies the lowest-lying state as the flavor singlet eta.  

\enlargethispage{\baselineskip}
The propagator and kernels are not intertwined in the quark-connected 
contribution. 
By inserting ${\bf 1}=\tau^2$ at the source and commuting one $\tau$ to the 
sink, the parity partners are made explicit, \vskip-2ex
\al{
 C^{\rm con}_{\Gamma_a,\Gamma_b}(n_0) &= \phantom{\big(} 
 \Braket{S^E \tau^2 \Gamma_a S^E \Gamma_b}+\Braket{S^O \tau^2 \Gamma_a S^O \Gamma_b} =
 \Braket{S^E \tau\Gamma_a S^E \Gamma_b\tau }-\Braket{S^O \tau\Gamma_a S^O \Gamma_b\tau}
 \nn\\
 &=\left( 
 \Braket{S^E \tau_0\Gamma_a S^E \Gamma_b\tau_0}-\Braket{S^O \tau_0\Gamma_a S^O \Gamma_b\tau_0} 
 \right)\times (-1)^{n_0}. 
}
Hence, the kernel $\Gamma$ also overlaps with states that would be na\"ively 
associated with a kernel $\Gamma\tau_0$. 
These extra states have opposite parity, which can be understood from the 
explicit form of $\tau_0$.

The extra states also have $m_3^2\to-m_3^2$ and change the sign of terms with two 
instances of $S^O$, which becomes clear by inserting the propagator with 
nested projections, \mbox{Eq.}~\eqref{eq:S}. 
Thus, these extra states are composed of unlike quark flavors in a 
taste-breaking combination. 
By either setting $\sfs\to-\sfs$ or $m_3\to-m_3$ on one of the two quark 
propagators, states with unlike quark flavors but without taste-breaking 
are obtained (for $\sfs\to-\sfs$ as extra states, for $m_3\to-m_3$ as 
conventional states). 
Note that there cannot be any quark-disconnected contribution to the extra 
states at all. 
Thus, $\gt=\Gamma_{a,b}\in\{\Gamma^{\omega O}\}$ yields exclusively charged 
pseudoscalars as extra states in the quark-connected contribution, since 
no quark-antiquark states with $J^{PC}=0^{+-}$ are realized in QCD. 
The quark-connected contribution for 
${\bf 1}=\Gamma_{a,b}\in\{\Gamma^{\eta E}\}$ also contains charged 
pseudoscalars as extra states, though it also contains the conventional 
scalar with a quark-disconnected contribution starting at $\mathcal O(1)$. 
These extra ground states are genuine charged pions due to the 
non-existence of a quark-disconnected contribution. 
Lastly, there are scalar extra states for 
$i\gt\gf=\Gamma_{a,b}\in\{\Gamma^{\eta E}\}$, which must be interpreted as 
charged scalars and are actually unphysical. \vskip-2ex

\vskip-1ex
\enlargethispage{\baselineskip}
\section{Conclusions}
\vskip-1ex

In these proceedings I have discussed QCD with flavored minimally doubled 
fermions. 
With a non-singlet mass term there is a natural representation of the 
SU(2) flavor group at finite lattice spacing. 
I have presented a proof that the quark determinant and thus also the QCD 
vacuum/medium are protected from the explicitly broken charge conjugation 
and reflection symmetries. 
Flavored mesons can be identified easily by means of appropriate projection 
operators and are indeed distinguishable through their quark mass dependence 
and their quark-disconnected contributions. 
Taste-breaking can be removed without need of fine tuning.  
Appropriately defined correlation functions are protected from the broken 
time reflection symmetry. 
Field configurations as well as appropriately defined meson 
correlation functions are automatically $\mathcal O(a)$ improved and the 
quark mass dependence is quadratic as it befits chiral fermions. 
A more comprehensive and more extended discussion of QCD with flavored MDF 
is in preparation. 
With this evidence in place, it is about time to start using minimally 
doubled fermions as sea quarks in lattice QCD. 

\end{document}